# Bioinspired polymer-incorporating self-lubricating and antifouling hydrogels

Weifeng Lin[1,2,†], Monika Kluzek[1,†], Nir Kampf[1], Yifeng Cao[1,3], and Jacob Klein[1,*]

[1] Department of Molecular Chemistry and Materials Science, Weizmann Institute of Science, Rehovot 76100, Israel

[2] Key Laboratory of Bio-Inspired Smart Interfacial Science and Technology of Ministry of Education, School of Chemistry, Beihang University, Beijing 100191, PR China

[3] Institute of Zhejiang University-Quzhou, Quzhou, Zhejiang 324000, PR China

† Co-authors: contributed equally to this work

*Corresponding author: jacob.klein@weizmann.ac.il

**Abstract**

Healthy articular cartilage has excellent lubricating properties, with friction coefficients reaching $\mu \approx$ 0.002 – 0.02 at physiological pressures. Such high-performing lubricating layer in joints is attributed to the surface hydration arising from the interplay between multiple hydrophilic biopolymers (such as hyaluronic acid, proteoglycans, and lubricin) and phospholipids in the cartilage matrix. Mimicking such molecular structure, hydrogels, composed of a hydrophilic polymer network, have the potential to replicate the lubricating feature and possibly replace natural cartilages.

In this study, we have synthesized a poly(2-methacryloyloxyethyl phosphorylcholine-*co*-*N*-isopropylacrylamide) (PMPC-*co*-PNIPAM, PMN) random copolymer with highly-hydrated lubricous 2-methacryloyloxyethyl phosphorylcholine moieties[1-3] and less hydrated *N*-isopropylacrylamide moieties. Incorporation of PMN copolymers within various hydrogels significantly reduces the gels' sliding surface friction, resulting in low friction coefficients against different counter surfaces, including stainless steel (hard metal surface), polyethylene (hydrophobic surface), and polyHEMA (soft hydrogel surface). Additionally, hydrogels containing PMN are shown to be biocompatible and have excellent antifouling properties, making them an ideal coating for commercially available stents. With these qualities, hydrogels containing PMN stand out as a promising new material with numerous possible applications.

1. **Introduction**

Natural articular cartilage is a thin layer of porous hydrated tissue with low friction that is unmatched by any artificial material.[4, 5] Damage caused by trauma or aging can lead to joint diseases, such as osteoarthritis.[6, 7] Hydrogels, which are widely used in biomedical and other applications,[8-10] are considered suitable replacement materials for damaged articular cartilage due to their good biocompatibility, liquid permeability, and wettability.[11, 12]

In recent years, several methods have been used to prepare lubricated hydrogels.[13-15] Recently, Milner et al.[16] incorporated a biomimetic boundary lubricant poly(2-methacryloyloxyethylphosphorylcholine) (PMPC) polymer network into a high-strength poly(2-acrylamido-2-methylpropanesulfonic acid)-polyacrylamide double network (DN) gel, forming a three-component network (PMPC-DN) hydrogel. Preliminary tribological studies show that the friction coefficient of both DN and PMPC-DN hydrogels decreases with increasing sliding speed, which is consistent with two-phase lubrication in the physiological sliding velocity range. the presence of the boundary-lubricated PMPC network, the friction coefficient of the PMPC-DN hydrogel is roughly half that of the DN hydrogel. Inspired by the unique structure of the articular cartilage, Rong et al.[17] reported a bilayer hydrogel material in which thick hydrophilic polyelectrolyte brushes on the topmost of the hydrogel provides effective water-based lubrication, while the hard hydrogel layer as the substrate provides load-bearing capacity. Their synergy enables a low friction coefficient under heavy load conditions in water, and its performance is comparable ~~close~~ to that of natural articular cartilage.

Inspired from the low friction of cartilage boundary lubrication, where phosphatidylcholine (PC) lipids whose highly hydrated phosphocholine headgroups may reduce friction via the hydration lubrication mechanism,[18-20] we incorporated trace lipid concentrations into hydrogels to create a molecularly thin, lipid-based boundary layer that renews continuously. We observed a 80% to 99.3% reduction in friction and wear relative to the lipid-free gel, over a wide range of conditions.[21] The reduction in friction and wear is attributed to a lipid-based boundary layer at the hydrogel surface, which is continually reconstructed as it wears, through progressive release of lipids. This effect persists when the gels are dried and then rehydrated. However, the reduction in friction is only observed when the counter surface is hydrophilic surface (such as glass, hydrogel, or stainless steel) but not with hydrophobic surface. This may be due to the low adsorption of lipids on the hydrophobic surface, so that the boundary layer would not have been well formed. This may impede the hydrogel in many applications where hydrophobic surfaces are implicated, such as artificial joints or plastic stents.

In this study, we synthesized a novel poly(2-methacryloyloxyethyl phosphorylcholine-co-*N*-isopropylacrylamide) (PMN) random copolymer (Fig. 1) and incorporated it at ca. 1 wt-% fraction within various hydrogels. In comparison to PMN-free hydrogel, we observed a strong reduction in friction in PMN-containing hydrogel when sliding against different counter surfaces, such as stainless steel (hard metal surface, partly hydrophilic), polyHEMA (soft hydrogel surface), and in particular polyethylene (a highly hydrophobic surface).-These highly lubricating PMN-hydrogels have good biocompatibility and antifouling qualities, allowing them to be used as a coating for commercially available stents.

## 2. Materials and methods

### 2.1 Materials

2-Methacryloyloxyethyl phosphorylcholine (MPC, 97%), *N*-isopropylacrylamide (NIPAM, ≥99%), ammonium persulfate (APS, 98%), *N,N,N',N'*-tetramethylethylenediamine (TEMED, 99%), 2-hydroxyethyl methacrylate (HEMA, 99%), ethylene glycol dimethacrylate (EGDMA, 98%), deuterium oxide (99.9 atom% D), Transwell® cell culture inserts (12 mm inserts, 0.4 μm pore size), XTT cell proliferation assay, Arabinose were purchased from Sigma-Aldrich (Israel). Ampicillin was purchased from Tivan Biotech (Israel). RPMI medium (Gibco), fetal bovine serum (FBS; Gibco), penicillin−streptomycin, Vybrant DiD cell-labeling solution, pBAD Kit (V43001) were purchased from Thermo Fisher Scientific Inc (Israel). Water used was purified using a Barnstead NanoPure system (ThermoFisher Scientific, Waltham, MA, USA) to MΩ.cm (at 25°C) and total organic content < 1 ppb.

### 2.2 Synthesis and characterization of a poly(2-methacryloyloxyethyl phosphorylcholine-*co*-N-isopropylacrylamide) random copolymer

A procedure for conventional copolymerization (Figure 1), to prepare the random copolymer with the content of the 12.5 mol % MPC unit, was performed as follows: MPC (3 g, 10.2 mmol), NIPAM (12 g, 106 mmol), and APS (320 mg, 1.4 mmol) were dissolved in 200 mL of pure water. The solution was deoxygenated by nitrogen for 30 min. After injecting with 44 μL TEMED, polymerization was carried out at room temperature for 16 h under a nitrogen atmosphere. The reaction mixture was dialyzed against pure water for 48 h. The random copolymer was obtained using a freeze-drying technique (14.7 g, 91.9%).

## 2.3 Hydrogel preparation

PMN-free pHEMA hydrogels were prepared as follows: HEMA (3 mL), EGDMA (0.1 mL or 0.3 mL), APS aqueous solution (0.2 mL, 24 mg APS), and purified water (2 mL) were added together and vigorously stirred for 15 min at room temperature. The molar ratio of the EGDMA cross-linker to the HEMA monomer was kept at 2 %. To accelerate the reaction, TMEDA (50 μL) was added to the mixture, which was then stirred for 20 seconds and poured into a 6 cm diameter petri dish. The mixture was allowed to gel at room temperature for 4 hours, followed by immersion in an excess of purified water for 3 days to remove unreacted materials. This resulted in very slight swelling of the pHEMA hydrogels, whose final water content was measured at 38.3 ± 0.3 % of the gel weight (determined by weight-changes upon freeze-drying). Obtained hydrogels were cut into circular disks of diameter 20 mm and thickness ~2 mm, and the surface exposed to air during the gelation was used for measurements.

Liposome-incorporating gels were prepared similarly save that the 2 mL purified water was replaced by 2 mL of the respective PMN solution at concentration 25 mg/mL.

## 2.4 Rheometry

Viscoelastic measurements were performed using a strain/stress-controlled rheometer (Thermo-Haake, Mars III, Karlsruhe, Germany). Samples were prepared in the form of disks with a diameter of 20 mm and thickness ~2 mm. Samples were tested using a parallel plate geometry. The samples were placed between the rheometer plates and a slight compression of ~1.0 N was applied. Amplitude sweep tests were performed in constant strain mode for strains in the range of $10^{-4}$ to 0.5, the oscillation frequency being 1 Hz. Frequency sweep tests were performed in stress control mode, with stress being in the range of linear viscoelasticity, as determined from the amplitude sweep studies. The frequency range was between 0.05 and 30-100 Hz.

## 2.5 Friction measurments

Friction measurements of several independent experiments (different hydrogel samples with at least 3 different contact points) for each set of conditions, were carried out using a CETR UMT tribometer (Bruker, MA, USA) with normal and friction force sensors under PBS at 37 °C. Measurements were done on hydrogel samples (20 mm disks) and carried out at different sliding velocities ($v_s$) and loads ($F_n$) and under high-purity water (unless stated differently). Loads used were in the range 1.5-30 N. $F_s$ was determined from its value in the plateau region of the friction traces (friction coefficients evaluated at each load as $\mu = F_s/F_n$).

## 2.6 Dynamic light scattering

pHEMA (20 kDa, w/o PMN) was dissolved in ethanol/DMF (v/v=1/1), and pippeted (~ 5 µL] into water with final concentration (pHEMA) of 0.2 mg/mL. Mixture was than dialysed against water for 2 days. Size distribution and zeta potential of the samples were characterized using a Zetasizer Nano-ZS instrument (Malvern Instruments Ltd., Malvern, WR, U.K.) equipped with a red laser (633 nm) and a scattering angle of 173°.

## 2.7 Cryo-SEM freeze fracture imaging

Hydrogel samples (20 mm disks) were sectioned to slices of thickness 100 μm using a vibratome (Electron Microscopy Sciences, USA). The slices were sandwiched between an aluminum disc with a depression of depth 150 μm and a flat disc (M. Wohlwend GmbH, Switzerland) and were then cryo-immobilized by high pressure freezing (HPM010, BalTech, Liechtenstein). The frozen sample was removed from the disc under liquid nitrogen, mounted perpendicularly in a holder, transferred to a BAF 60 freeze fracture device (Leica Microsystems, Austria) using a VCT 100 cryo transfer shuttle (Leica) and was fractured perpendicularly to the plane of the frozen slice at -120 °C under a pressure of about $5\times10^{-7}$ mbar. The cold fractured surface was sometimes "etched" by increasing the temperature to about -105 °C for several minutes to let some frozen water sublime. The fractured sample was then transferred to an Ultra 55 cryo-scanning electron microscope (Zeiss, Germany) and observed using an InLens Secondary electrons detector at an acceleration voltage ranging between 1-2.5 kV.

## 2.8 Confocal fluorescence microscopy imaging

To prepare PMN-based hydrogels for confocal fluorescence microscopy imaging, hydrogels were prepared as above with modificaltion: the 2 mL purified water was replaced by 2 mL PMN with 0.5 wt% of acryloxyethyl thiocarbamoyl rhodamine B. Confocal pictures were acquired using confocal laser scanning microscopy LSM700 (Zeiss, Germany). All images were acquired using a 40X oil immersion objective (NA 1.4 Zeiss, Germany) and recorded in brightfield mode and in confocal mode using a 540 nm excitation laser channel, with a 0.3 μm optical slice step for z scanning. Picture analysis was performed using ImageJ software v1.52i (NIH, USA).

### 2.9 Commercialized catheter coating

Actreen® Hi-Lite Cath Nelaton (the catheter for male intermittent self-catheterization) was washed with excess water to remove the existing lubricant. The cleaned catheter was plasma ($O_2$/Ar=1/1) treated for 5 mins, and then immersed into 3% 3-(trimethoxysilyl) propyl methacrylate (toluene solution) for 12 h. The catheter was washed with toluene, and dried in the air. Hydrogel coating (w/o PMN) on the catheter was performed by pulling out the catheter from the polymerization solution after 1 min, and the polymerization was proceeding for 12 h under $N_2$ atmosphere.

### 2.10 Cell culture

Vero E6 cells were cultured in RPMI medium, supplemented with 1% Pen/Strep and 10% fetal bovine serum (FBS), and were maintained at 37 °C in a humidified atmosphere of 95% air and 5% $CO_2$.

### 2.11 Cytotoxicity studies

*Cell culture*

Prior cytotoxicity experiment, cells were cultured in 12-well plates at $8 \times 10^4$ cells per mL and allowed to attach for 24 h.

*Cytotoxicity evaluation*

The cytotoxicity of hydrogel containing 3 mg/mL PMN was conducted in 12-well plates using Transwell cell culture inserts. Hydrogel (0.6 mm disk) was placed in a center of Transwell and preincubated in a cell culture media for 1 h at room temperature in order to exchange water within the hydrogel. The inserts with hydrogels were then placed into wells containing the cultured cells and incubated for 48 h at 37 °C (95% air and 5% $CO_2$).

Cell viability was determined by the production of the yellow formazan product upon cleavage of XTT by mitochondrial dehydrogenases in viable Vero cells. After 48 h incubation with hydrogels, the cells were incubated with 500 μL of XTT solution for 4 h at 37 °C (95% air and 5% $CO_2$). Absorbance values were later measured with ClarioStar microplate reader (BMG LABTECH GmbH, Germany) at a wavelength of 450 nm. Background absorbance was measured at 620 nm and subtracted from the 450 nm measurement. The absorbance of a solution in RPMI-cell-free media with XTT was subtracted from all samples. The potential toxic effect of the different liposomal formulations tested was expressed as a viability percentage calculated using the following formula:

%Viability= [(ODtest/ODc)×100]

Where ODtest was the optical density of those wells with hydrogels, and ODc was the optical density of cells in RPMI media only (no hydrogel).

### 2.12 *In-vitro* PMN release studies

PMN (Rhodamine-conjugate) hydrogels were placed in 2 mL PBS (pH 7.2, osmol=290 mOsmo/kg) in 12-well plates and incubated at 37 °C (95% air and 5% $CO_2$) to mimic physiological conditions. At specific time points, 2 mL of PBS were taken and replaced with fresh buffer. The retrieved conditioned PBS was evaporated using a lyophilizer followed by resuspension in 0.2 mL PBS. The concentration of PMN-rhodamine-conjugate in PBS was determined with a ClarioStar microplate reader (BMG LABTECH GmbH, Germany) at a wavelength of 566 nm. PMN-rhodamine-conjugate stock solution used for hydrogel preparation was diluted to specific concentrations and used as a standard calibration curve. The percentage of the cumulative amount of released PMN was calculated from a standard calibration curve.

### 2.13 Fluorescence-based monitoring of gel-to-stainless steel (SS) and gel-to-polyethylene (PE) transfer of PMN- rhodamine conjugate during sliding

PMN-rhodamine conjugated based hydrogels, were prepared as above with small modification: 2 mL purified water was replaced by 2 mL PMN with 0.5 wt% of acryloxyethyl thiocarbamoyl rhodamine B. PMN transfer *via* SS-gel (stainless steel, SS) or PE-gel (polyethylene, PE) sliding and imaging of transferred material. The spherical head (SS or PE) was slid past hydrogels incorporating PMN-rhodamine-conjugate or DMPC-MLV stained with 1% DiI, at a sliding velocity 1 mm/s for 5 min. It was then placed in a petri dish and the area of absorbed polymer or lipids transferred during the sliding was imaged by excitation of rhodmamine and DiI dye at 532 nm using a Typhoon FLA 9500 scanner (GE Healthcare Bio-Sciences AB, Sweden), with photomultiplier tube set up at 500´ gain and pixel size 50 μm. Pictures were analyzed using ImageJ software (NIH, USA). Subsequently, the spherical head (PE or SS) head was washed with water (PMN) or ethanol (DMPC), and evaporated overnight using a nitrogen stream followed by lyophilization. 400 μL of water (PMN) or chloroform (DMPC) was used to re-suspend polymer or lipids, and the solution was placed in a quartz silica cuvette with a 1 mm path length. Acquisition of rhodamine and DiI emission spectra was performed with an Agilent Cary Eclipse Fluorescence Spectrophotometer (Varian Instruments, Walnut Creek, CA) at room temperature (Figure S1). The excitation wavelength was set at 500 nm with a bandpass of 20 nm, and the emission was also recorded with bandpass of 20 nm. Spectrum acquisition was repeated over a minimum of 3 separate samples. Calibration curves were prepared by measuring the maximal (Insert of Figure 5 B and C) fluorescence intensity of known amounts of pure PMN-Rhodamine-water and DMPC-DiI-chloroform solution by the same method and parameters to ensure the same experimental conditions.

**2.14 Anti-biofouling measurements**

*Vero E6 cells adhesion to hydrogels*

Vero E6 cell suspension (1.5 mL of $5.2 \times 10^5$) was seeded onto 6 mm hydrogel discs (neat and PMN) in 24 well plates and allowed to attach for 24 h. The cells were then stained with Vybrant DiD by adding 5 μL of stock solution and incubated for 20 min at 37 °C. After 20 min hydrogels were washed carefully with PBS and placed in a fresh cell medium. Hydrogels were placed on the slide and observed with confocal laser scanning microscopy LSM700 (Zeiss, Germany). All images were acquired using a 60X oil immersion objective and recorded in brightfield mode and in confocal mode using a 635 nm excitation laser channel. The anti-biofouling property of hydrogel was further determined by using Typhoon FLA 9500 scanner (GE Healthcare Bio-Sciences AB, Sweden) with 635 nm laser wavelength, photomultiplier tube set up at 250´ gain and pixel size 25 μm. Pictures were analyzed using ImageJ software (NIH, USA). For comparative analysis, all parameters during image acquisition were kept constant throughout each experiment. Additionally, fluorescence signal on hydrogel surfaces was determined with a ClarioStar microplate reader (BMG LABTECH GmbH, Germany) in 96-well plate at a wavelength of 630 nm with top reading and spiral scan mode [Scan diameter: 6 (mm), No. of flashes per well 106].

*Bacteria adhesion to hydrogels*

GFP-expressing *E. Coli* culture was developed by plasmid transfection. The *sf-gfp-his* plasmid was cloned into a pBAD24. The plasmid was then transformed into *E. coli* MG1655 imp4213. *E. coli* (MG1655) precultures cultures were inoculated at OD600 0.05 from an overnight culture, and growth was carried out at 37 ˚C with shaking for 4 h, in Luria-Bertani broth (LB) supplemented with ampicillin (100 mg/L) and 0.2% arabinose until mid-exponential phase (OD at 600 nm ~ 0.5). Then bacteria were diluted in BM2G minimal medium [62 mM potassium phosphate buffer, pH 7, 7 mM (NH4)2SO4, 2 mM MgSO4, 10 μM FeSO4, 0.4% (wt/vol) glucose, 1% Monosodium glutamate] to reach OD 0.08. Hydrogel discs (diameter 6 mm) were placed in 24-well plate and 1 mL of bacteria suspension was added and incubated at 37 °C without shaking for 12 h. Subsequently, hydrogels were washed carefully with PBS and placed in fresh PBS. Hydrogels were placed on the slide and observed with confocal laser scanning microscopy LSM700 (Zeiss, Germany). All images were acquired using a 60X oil immersion objective and recorded in brightfield mode and in confocal mode using a 473 nm excitation laser channel. Moreover, hydrogels were further visualized using Typhoon FLA 9500 scanner (GE Healthcare Bio-Sciences AB, Sweden) with 635 nm laser wavelength, photomultiplier tube set up at 250´ gain, and pixel size 25 μm. Pictures were analyzed using ImageJ software (NIH, USA). For comparative analysis, all parameters during image acquisition were kept constant throughout each experiment. Additionally, fluorescence signal on hydrogel surfaces was determined with a ClarioStar microplate reader (BMG LABTECH GmbH, Germany) in 96-well plate at a wavelength of 470 nm with top reading and spiral scan mode [Scan diameter: 6 (mm), No. of flashes per well 106].

## 3. Results and Discussion

*Synthesis and structural characterization of PMN co-polymer*

PMPC-co-PNIPAM (PMN) random co-polymer was successfully synthesized via conventional polymerization method and its structure was confirmed by $^1$H NMR analysis (Figure 1). The characteristic peak for NIPAM methine protons is at $\delta$ = 3.9 ppm, while MPC pendant methylene protons shows peak at $\delta$ = 3.7 ppm. The estimated 7:1 ratio between NIPAM and MPC was determined by comparing the integral intensity signal. The number-average molecular weight ($M_n$) and the molecular weight distribution ($M_w/M_n$) estimated from gel-permeation chromatography (GPC) were $2.2 \times 10^5$ and 1.64, respectively. It can be calculated that the units of MPC and NIPAM are ~200 and 1400.

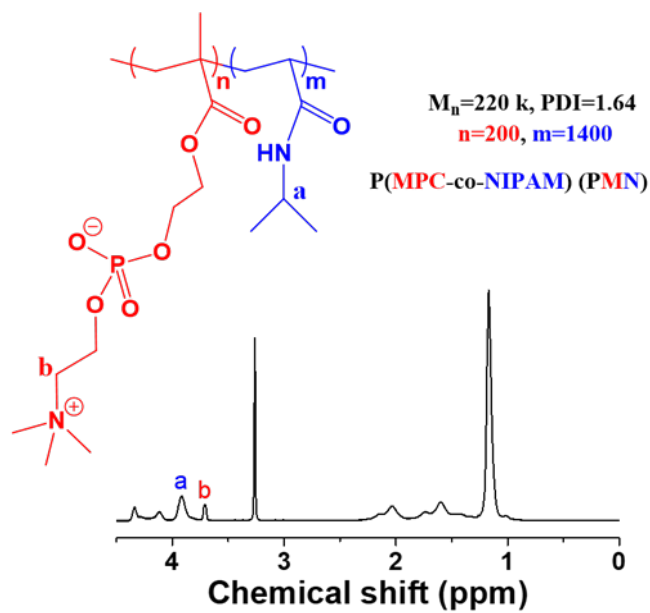

**Figure 1.** Chemical structure and ¹H-NMR spectra of poly(2-methacryloyloxyethyl phosphorylcholine-co-*N*-isopropylacrylamide) (PMN) co-polymer in $D_2O$.

*Evaluation of the lubrication property of PMN-containing hydrogels*

PMN co-polymer was incorporated into biocompatible pHEMA hydrogels[22] by mixing a low concentration (1% wt.) of PMN with the desired HEMA solution. Following polymerization and cross-linking, the resulting pHEMA hydrogel shows formation of local PMN-microreservoirs dispersed throughout the hydrogel bulk. Freeze-fracture cryo-scanning electron microscopy (cryo-SEM) revealed that PMN-pockets size ranges from 200 nm up to 900 nm (Figure 2 A) with homogenous distribution (Figure S2), which was further confirmed by confocal microscopy imaging (Figure 2 B), while PMN-free hydrogels show featureless internal structure (Figure 2 C). Under physiological conditions, PMN pockets within pHEMA hydrogels are stable long-term, with less than 1% PMN leakage from the cavities over 18 days (Figure S3).

Moreover, the incorporation of PMN alters the hydrogel morphology, as visualized via confocal microscopy. Neat (PMN-free) pHEMA hydrogels show a porous structure at the surface with cavities ranging from 3 µm up to 15 µm in size (Figure S1), while PMN-incorporating hydrogels display a smooth and cavity-free surface. These morphological differences are also visible in the opacity of PMN-pHEMA hydrogels in contrast with the clear neat pHEMA materials.

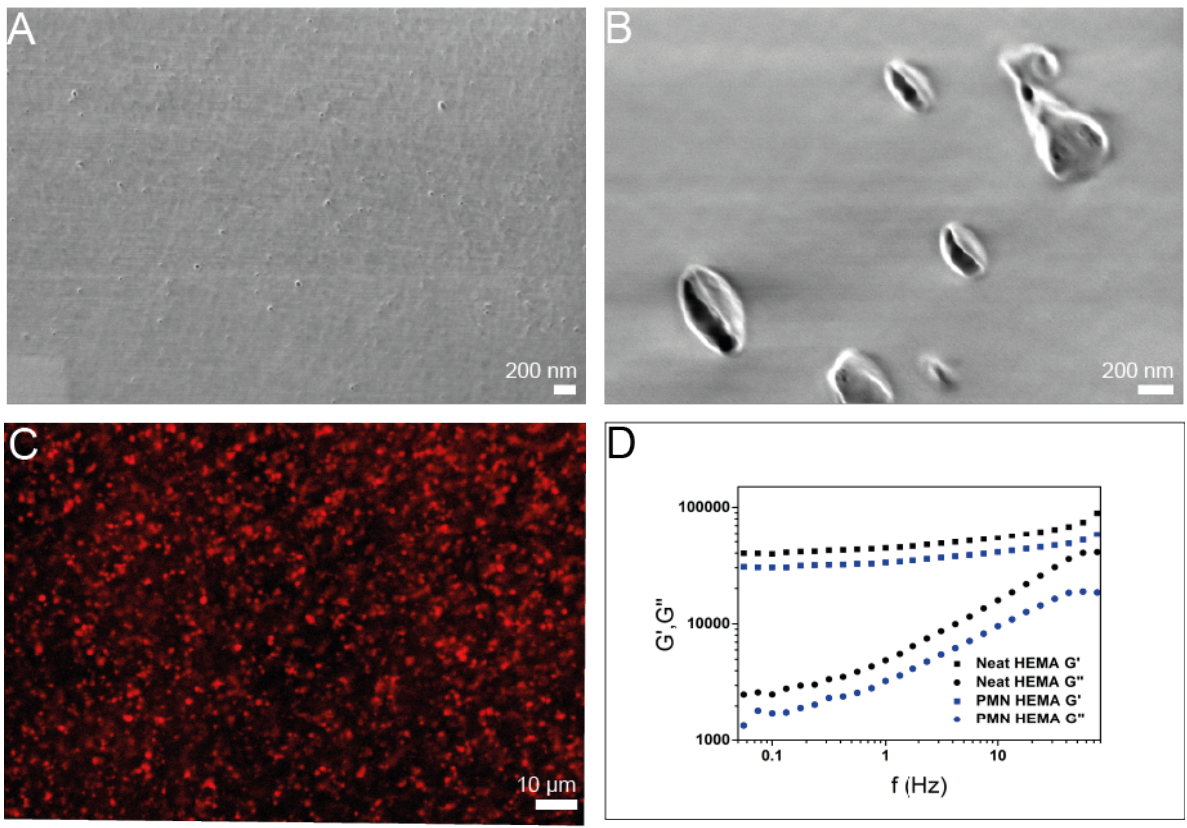

**Figure 2.** Characterization of PMN- free and PMN-incorporated hydrogels. Freeze-fracture surface of PMN-free (**A**) and PMN-incorporated hydrogel. (**C**) Confocal microscopy section of the hydrogel incorporating fluorescently labeled PMN pockets recorded at RT. (**D**) Storage and loss moduli of PMN-free and PMN-incorporating (2%) pHEMA hydrogels.

Rheometrically determined mechanical properties reveal that the hydrogel storage modulus G' (>>G", the loss modulus) varies, over a frequency (f) range of 0.1 to 10 Hz, by ~50% or less between PMN-free and PMN-based hydrogels for 2% cross-linking hydrogel (Figure 2 D).

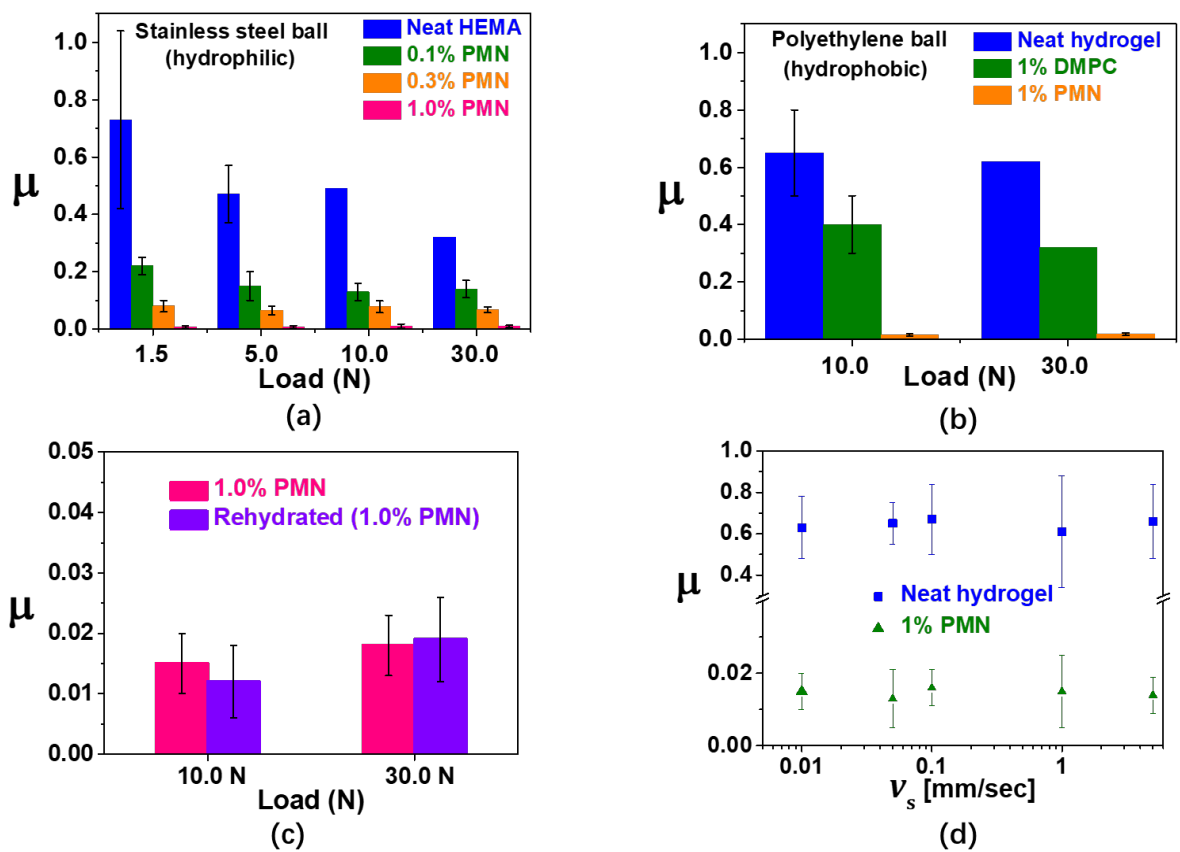

**Figure 3.** (**A**) Variation of the friction coefficient, μ, with different concentration of PMN (from 0 to 1%) incorporated in the hydrogel, at different loads between a pHEMA gel and a sliding steel sphere. (**B**) Variation of the friction coefficient, μ, with different materials in the hydrogels at different loads between a pHEMA gel and a polyethylene sphere. (**C**) Effect of rehydration after dehydration of PMN-free and PMN-incorporating gels, showing retention of the characteristic self-lubricating ability for the latter. (**D**) Variation of the coefficient of friction μ with sliding velocity vs for PMN-free (blue) and PMN-incorporating (green) pHEMA gels. Error bars indicate SD from at least three measurements.

The sliding friction $F_s$ between the hydrogel and a polished stainless-steel surface, as well as surfaces of other materials, was examined over a range of loads $F_n$, yielding the coefficient of sliding friction $\mu = F_s/F_n$. A reduction in friction of coefficient is seen for the PMN-based hydrogels relative to the PMN-free hydrogels (Figure 3 A). For 1% incorporated PMN, the reduction in friction of coefficient μ ranges from 95% to 99% at different loads. The reduction could be applied with other hydrogels (Figure S4). Compared to the control groups (different homopolymers, in Figure S5), PMN showed the strong synergistic effect. In order to know the interaction between pHEMA and PMN, size and zeta potential measurements were performed (Figure S6). pHEMA is not soluble in water (there was precipitation after dialysis with pHEMA). The size distribution of pHEMA/PMN dispersion shows one peak without precipitation, which could be attributed to the interaction between pHEMA and NIPAM moiety, and the hydrated PMPC moieties are exposed.

The counter surfaces were extended from hydrophilic surface to hydrophobic surface for more potential application. While lipid-based hydrogels show almost little improved lubrication (about 10-20% less friction of coefficient), PMN-based hydrogels display more than 95% in friction reduction (Figure 3 B). When the PMN-incorporating hydrogel is completely dried and then rehydrated, friction returns to low values and lubrication is self-sustaining (Figure 3 C). Finally, the incorporated PMN reduces friction as well as wear and surface damage (Figure 4). After 1 hour of sliding, the appearance of the PMN-incorporated gel was unchanged, whereas the PMN-free gel and lipid-incorporated gel was torn after 5 mins of back-and-forth shear under the load. Furthermore, a hydrophobic commercialized catheter was used to compare the lubrication between our hydrogel coating and the commercial lubricant coating (glycerol). The PMN-based hydrogel coating shows 80-90% reduction in friction (with polyethylene sphere) compared to the commercial lubricant group (fig. S8). Finally, and similar to the low friction lipid-incorporating hydrogels[21], the PMN-icorporating hydrogels could be fully dehydrated and then fully recovered their lubricating ability following rehydration (fig. S9).

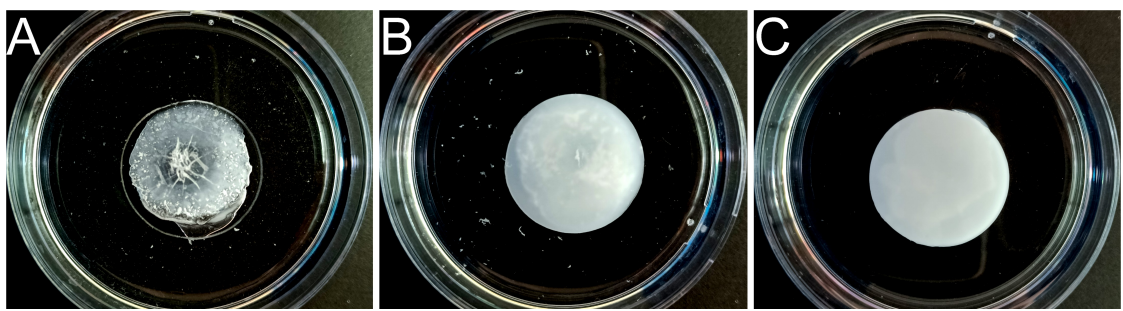

**Figure 4**. Appearance of gel samples under 10-N load on the polyethylene sphere. After 1 hour of sliding, the appearance of the (**A**) PMN-free hydrogel, and hydrogel incorporating (**B**) DMPC-lipid; (**C**) PMN-copolymer. Hydrogels are immersed in 3 mL of water. Lipid-incorporated gel was torn after 5 mins of back-and-forth shear under the load.

We interpret the observed differences in friction between lipid incorporating- and PMN-incorporating hydrogels rubbed by a polyethylene countersurface to the poor transfer of lipids towards a hydrophobic surface upon sliding. Indeed, comparison of fluorescence signal measured on PE surfaces upon rubbing (Figure 5 A) shows a uniform signal coverage for PMN-hydrogels, indicating efficient transfer of the fluorescently labelled PMN polymer to the surface. Contrariwise, fluorescence signal on PE surfaces upon rubbing with DMPC-hydrogels shows weak signal with disrobed fragments of gel, in good agreement with the low-friction data.

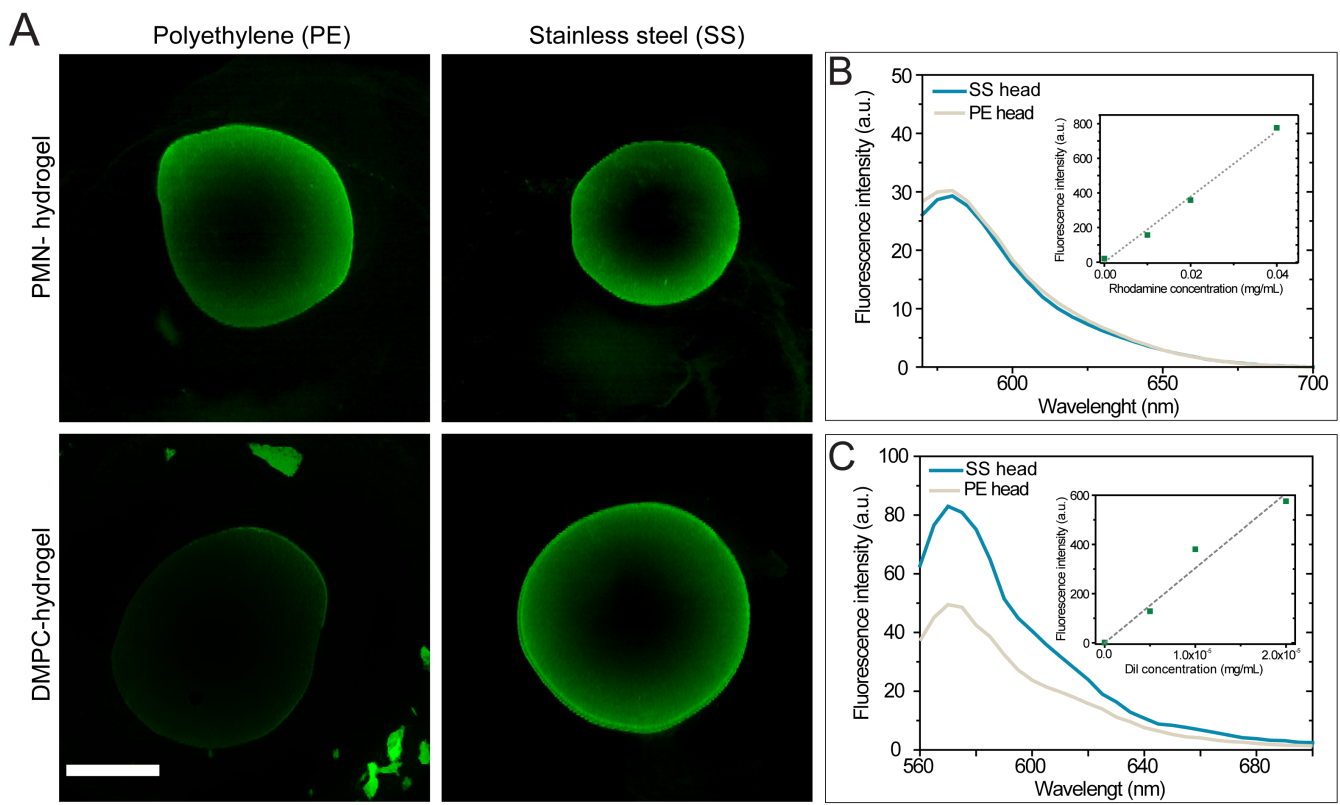

**Figure 5.** (**A**) Fluorescence signal recorded on PE and stainless-steel spherical ball upon 5 min sliding on rhodamine-labeled PMN and DiI-labelled liposomes incorporated in pHEMA hydrogels. Scale bar, 0.2 cm. Fluorescence signal of (**B**) PMN-rhodamine and (**C**) DMPC-DiI obtained from SS and PE heads upon 5 min sliding. Inserts: calibration curve of (**B**) Rhodamine-PMN in water, and (**C**) DMPC-DiI in chloroform solutions.

Replacing the countersurface from hydrophobic PE to the hydrophilic stainless steel (SS) shows uniform fluorescence intensity left on the rubbing surface for both PMN- and DMPC- hydrogels, suggesting efficient transfer for both formulations. We furthermore evaluated the amount of transferred material to the ball from fluorescence intensity calibration curves (Figure 5 B and C). Quantifications show that PMN-hydrogels deposit a similar amount of rhodamine-labelled PMN both in PE and stainless-steel surfaces upon rubbing (Table S1). DMPC-hydrogels show deposition of DiI-labelled DMPC towards stainless surfaces, in agreement with what reported by Lin et al.[21] whereas rubbing against PE ball yielded 5 times lower amount of transferred lipids. However, in case of DMPC-hydrogels rubbing against PE surfaces we observe a significant variability in the estimated deposited amount, possibly due to the presence of fluorescently labelled gel fragments (as visible in Fig. 5 A) affecting the overall fluorescence intensity of the recovered material.

These results indicate that PMN-hydrogel is significantly less affected by the surface hydrophobicity/hydrophilicity of it sliding countersurface, as it performs similarly in terms of lubrication and material deposition with two drastically different materials (PE and SS).

This feature makes PMN-hydrogel more flexible towards multiple applications compared to lipid-incorporating hydrogels for lubrication, which are, as we showed, limited in their lubricative properties towards hydrophobic surfaces.

*Biocompatibility and antifouling properties of PMN-hydrogel*

We determined the hydrogel cytotoxicity by exposing monolayers of Vero cells to either a PMN-free or a PMN-incorporating pHEMA hydrogels for 48 hours in a Transwell system (Figure 6 A and B). As shown in Figure 6 C, both pHEMA-only and PMN-loaded hydrogels had no negative impacts on cell viability, with >80% cell survival upon exposure to hydrogels indicating low cytotoxicity (according to ISO Standard 109993-5). These results point to PMN-hydrogels as excellent biomaterials for healthcare applications.

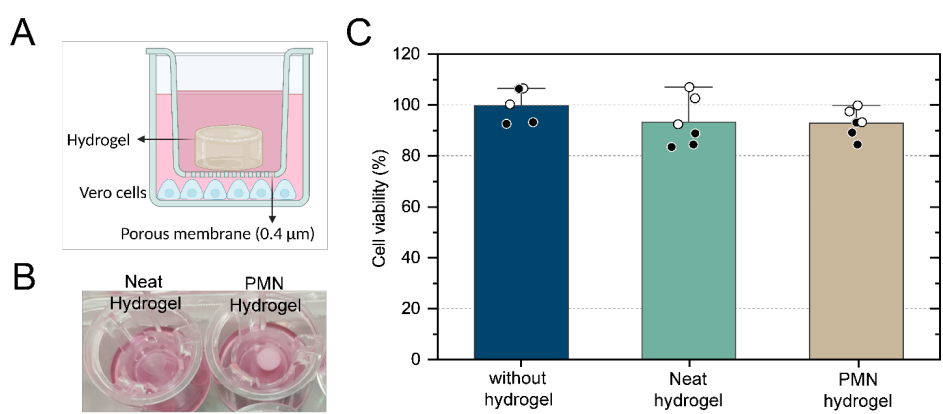

**Figure 6**. Cytotoxicity evaluation of PMN hydrogel (3 mg/mL PMN [is it 1%?]) on Vero cells. (**A**) Graphical illustration showing cells culture setup for evaluation of the cytotoxic effect of PMN-hydrogel. (**B**) Image of neat and PMN- hydrogels placed in Transwell (**C**) Cellular viability after 48 h incubation with hydrogels. Data represents two biological repeats (black and white data point) with a minimum of two technical repeats each.

We then evaluated the antifouling properties of PMN-hydrogels by probing both anti cell-adhesion and the bacteria adsorption towards the gels (Figure 7). We incubated either neat, PMN-incorporating, or DMPC-incorporating pHEMA hydrogels with fluorescently labelled Vero cells or GFP-expressing *E. Coli* bacteria suspensions for 24 hours and 12 h respectively. The hydrogels were then gently washed, examined with a fluorescence scanner (Figure 7 A and C), and quantified using a plate reader (Figure 7 B and D), as well as visualized *via* confocal microscopy (Figure S7).

Following incubation with Vero cells, we observe for neat pHEMA hydrogels a dense population of living cells 24 hours after incubation (Fig. 7 A and B), as expected for such gels due to the rough surface that favors cell attachment [ref]. For both DMPC- and PMN-containing hydrogels we instead observe no presence of cells in confocal microscopy (Fig. S7) and a significantly lower fluorescence intensity compared to pHEMA only gels, with fluorescence intensity ≥ 1000 a.u.

These results indicate that the presence of either lipids or PMN prevents cell adhesion on the gels' surfaces and provides good antifouling properties. Since in cell culture assays PMN-hydrogels showed no biocidal action (Figure 6 C), we ascribe the observed resistance to cell attachment to gel surface is purely due to its antifouling characteristic.

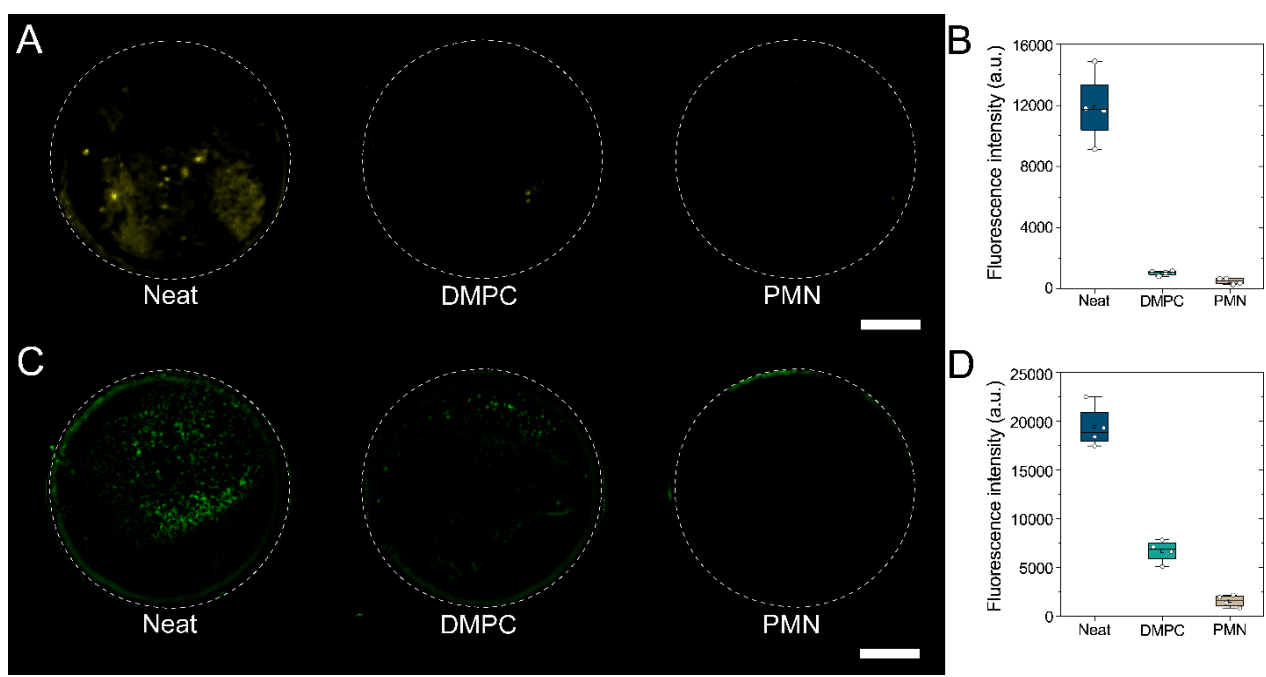

**Figure 7.** Typhoon-scanned fluorescence images of neat, DMPC, and PMN hydrogels and relative fluorescence intensity. (**A**) Representative images of hydrogels incubated for 24 h with Vero E6 cells labeled with Vybrant DiD (yellow). (**B**) Quantification of DiD fluorescence intensity of hydrogels obtained from plate reader measurements. (**C**) Representative images of hydrogels incubated for 12 h with GFP-expressing *E. coli* bacteria. (**D**) Quantification of GFP fluorescence intensity of neat, DMPC, and PMN hydrogels obtained from plate reader measurements. Fluorescence intensity measurements were conducted on 4 biological samples. Boxes represent the 25-75 percentiles of the sample distribution, with black vertical lines representing 1.5×IQR (interquartile range). Black horizontal line represents the median. Scale bar, 1.5 mm

In case of incubation with GFP-expressing bacteria, we observe that the fluorescence signal of PMN-hydrogel is 17 orders of magnitude lower compared to neat hydrogels and around 5 times lower than DMPC-hydrogels (Figure 7 D). Particularly, we observe for PMN-hydrogels that the strongest fluorescence signal arises from the borders (where the hydrogel was cut), whereas in neat hydrogels bacteria are present in the middle of the gel (Figure 7 C), suggesting that no bacteria are actually present on the PMN gel surface. Since both approaches utilized to quantify bacteria fluorescent signals rely on surface-related measurements, we cannot exclude that bacteria could be present within the bulk of neat pHEMA gels, due to their highly porous mesh (see Figure S1).

Both results with Vero cells and *E. Coli* corroborate that the lubricious nature of PMN-incorporating hydrogels provides excellent antifouling properties, with antibacterial features even higher than the equally lubricious DMPC-hydrogels.

## 4. Conclusion

Incorporation of PMN copolymers within hydrogels significantly reduces the gels' sliding surface friction, resulting in low friction coefficients against different counter surfaces, including stainless steel (hard metal surface), polyethylene (hydrophobic surface), and polyHEMA (soft hydrogel surface). In addition, the hydrogels exhibit good anti-bacterial adhesion and anti-cell adsorption. Such hydrogels reported here provides a new strategy for various biomedical applications.

## References

1. Ishihara, K., Highly lubricated polymer interfaces for advanced artificial hip joints through biomimetic design. *Polymer Journal* **2015,** *47* (9), 585-597.
2. Chen, M.; Briscoe, W. H.; Armes, S. P.; Klein, J., Lubrication at Physiological Pressures by Polyzwitterionic Brushes. *Science* **2009,** *323* (5922), 1698-1701.
3. Xie, R.; Yao, H.; Mao, A. S.; Zhu, Y.; Qi, D.; Jia, Y.; Gao, M.; Chen, Y.; Wang, L.; Wang, D.-A.; Wang, K.; Liu, S.; Ren, L.; Mao, C., Biomimetic cartilage-lubricating polymers regenerate cartilage in rats with early osteoarthritis. *Nature Biomedical Engineering* **2021,** *5* (10), 1189-1201.
4. Jahn, S.; Seror, J.; Klein, J., Lubrication of Articular Cartilage. *Annual Review of Biomedical Engineering* **2016,** *18* (1), 235-258.
5. Lin, W.; Klein, J., Recent Progress in Cartilage Lubrication. *Advanced Materials* **2021,** *33* (18), 2005513.
6. Lotz, M. K., New developments in osteoarthritis: Posttraumatic osteoarthritis: pathogenesis and pharmacological treatment options. *Arthritis Research & Therapy* **2010,** *12* (3), 211.
7. Loeser, R. F., Aging and osteoarthritis. *Current Opinion in Rheumatology* **2011,** *23* (5), 492-496.

8. Green, J. J.; Elisseeff, J. H., Mimicking biological functionality with polymers for biomedical applications. *Nature* **2016,** *540* (7633), 386-394.
9. Peppas, N. A.; Hilt, J. Z.; Khademhosseini, A.; Langer, R., Hydrogels in biology and medicine: from molecular principles to bionanotechnology. *Advanced materials* **2006,** *18* (11), 1345-1360.
10. Zhang, Y.; Xu, R.; Zhao, W.; Zhao, X.; Zhang, L.; Wang, R.; Ma, Z.; Sheng, W.; Yu, B.; Ma, S.; Zhou, F., Successive Redox-Reaction-Triggered Interface Radical Polymerization for Growing Hydrogel Coatings on Diverse Substrates. *Angewandte Chemie International Edition* **2022,** *61* (39), e202209741.
11. Ngadimin, K. D.; Stokes, A.; Gentile, P.; Ferreira, A. M., Biomimetic hydrogels designed for cartilage tissue engineering. *Biomaterials Science* **2021,** *9* (12), 4246-4259.
12. Zhou, L.; Guo, P.; D'Este, M.; Tong, W.; Xu, J.; Yao, H.; Stoddart, M. J.; van Osch, G. J. V. M.; Ho, K. K.-W.; Li, Z.; Qin, L., Functionalized Hydrogels for Articular Cartilage Tissue Engineering. *Engineering* **2022,** *13*, 71-90.
13. Shoaib, T.; Espinosa-Marzal, R. M., Advances in Understanding Hydrogel Lubrication. *Colloids and Interfaces* **2020,** *4* (4), 54.
14. Lin, W.; Klein, J., Hydration Lubrication in Biomedical Applications: From Cartilage to Hydrogels. *Accounts of Materials Research* **2022,** *3* (2), 213-223.
15. Zhang, X.; Wang, J.; Jin, H.; Wang, S.; Song, W., Bioinspired Supramolecular Lubricating Hydrogel Induced by Shear Force. *Journal of the American Chemical Society* **2018,** *140* (9), 3186-3189.
16. Milner, P. E.; Parkes, M.; Puetzer, J. L.; Chapman, R.; Stevens, M. M.; Cann, P.; Jeffers, J. R. T., A low friction, biphasic and boundary lubricating hydrogel for cartilage replacement. *Acta Biomaterialia* **2018,** *65*, 102-111.
17. Rong, M.; Liu, H.; Scaraggi, M.; Bai, Y.; Bao, L.; Ma, S.; Ma, Z.; Cai, M.; Dini, D.; Zhou, F., High Lubricity Meets Load Capacity: Cartilage Mimicking Bilayer Structure by Brushing Up Stiff Hydrogels from Subsurface. *Advanced Functional Materials* **2020,** *30* (39), 2004062.
18. Goldberg, R.; Schroeder, A.; Silbert, G.; Turjeman, K.; Barenholz, Y.; Klein, J., Boundary lubricants with exceptionally low friction coefficients based on 2D close‐packed phosphatidylcholine liposomes. *Advanced Materials* **2011,** *23* (31), 3517-3521.
19. Seror, J.; Zhu, L.; Goldberg, R.; Day, A. J.; Klein, J., Supramolecular synergy in the boundary lubrication of synovial joints. *Nature communications* **2015,** *6* (1), 1-7.
20. Schmidt, T. A., Lubricating lipids in hydrogels. *Science* **2020,** *370* (6514), 288-289.
21. Lin, W.; Kluzek, M.; Iuster, N.; Shimoni, E.; Kampf, N.; Goldberg, R.; Klein, J., Cartilage-inspired, lipid-based boundary-lubricated hydrogels. *Science* **2020,** *370* (6514), 335-338.
22. Nicolson, P. C.; Vogt, J., Soft contact lens polymers: an evolution. *Biomaterials* **2001,** *22* (24), 3273-3283.

**Acknowledgments**

We would like to thank Prof. Avi Domb from Hebrew University of Jerusalem for the GPC measurements. We also wish to thank Dr. Asaf Levy (The Hebrew University of Jerusalem) for the bacteria strain, and Dr. Yaara Oppenheimer-Shaanan (The Hebrew University of Jerusalem, Weizmann Institute of Science) for scientific support in establishing an intimal bacterial cell line. We are grateful to Dr. Yoav Barak from the Department of Chemical Research Support (BioNANO lab) for microplate reader measurement assistance (Weizmann Institute of Science). Graphical schemes were created using Biorender platform (www.biorender.com)

We acknowledge with thanks the European Research Council (ERC AdG Cartilube, grant no. 743016), the Israel Ministry of Science (grant no. 315716), the Israel Science Foundation (grant ISF 1229/20), the Israel

Science Foundation−Natural Science Foundation China Joint Program (grants ISF-NSFC and 2577/17), and the McCutchen Foundation, for supporting some of the work described in this Account. This work was made possible in part by the historic generosity of the Harold Perlman family.

# SUPPORTING INFORMATION FOR

# Bioinspired polymer-incorporated self-lubricating and antifouling hydrogels

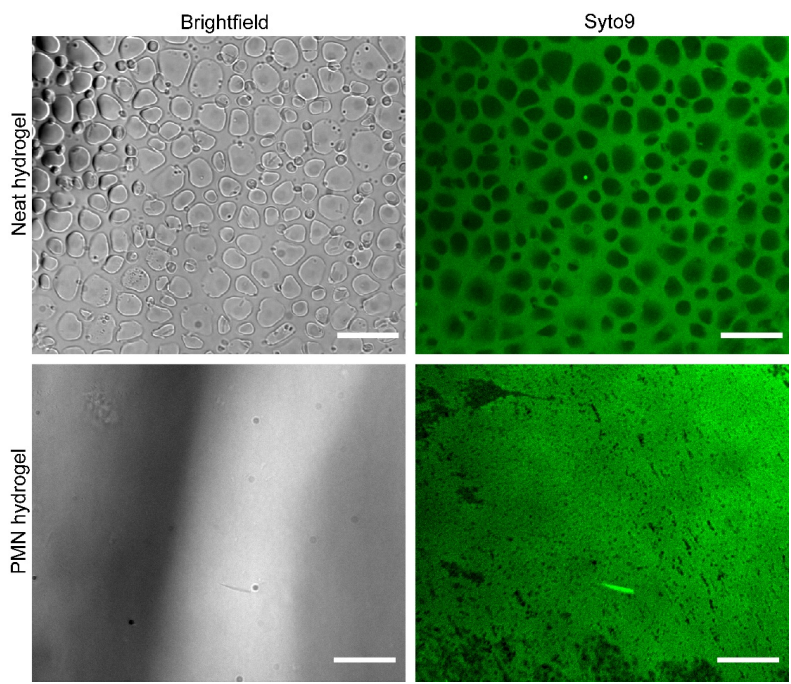

Figure S1. Brightfield (left column) and Syto9 (green, right column) fluorescence confocal microscopy images of neat and PMN-containing pHEMA hydrogels. Scale bar, 20 µm.

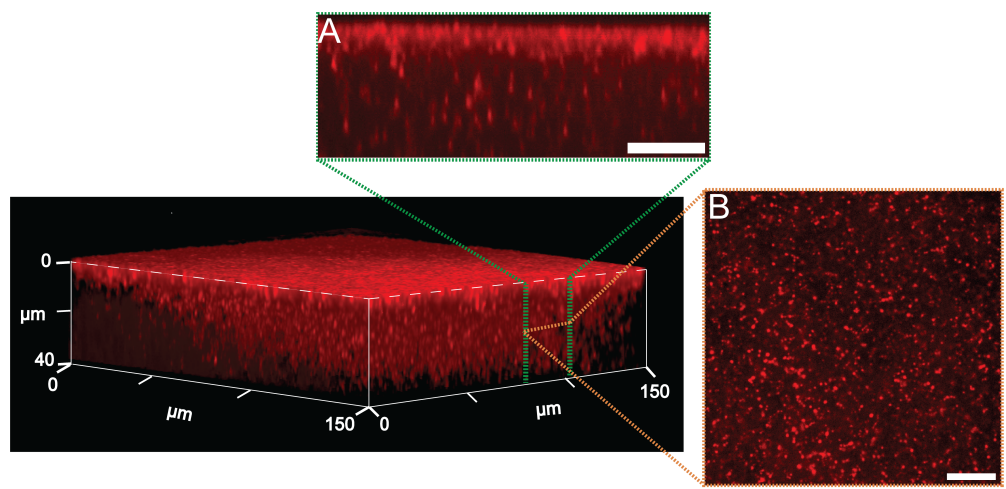

Figure S2. Confocal microscopy image showing sections through 40 µm thick slice of PMN-hydrogel. (A) shows orthogonal view and (B) in-plane view on hydrogel slice. Scale bar, 20 µm.

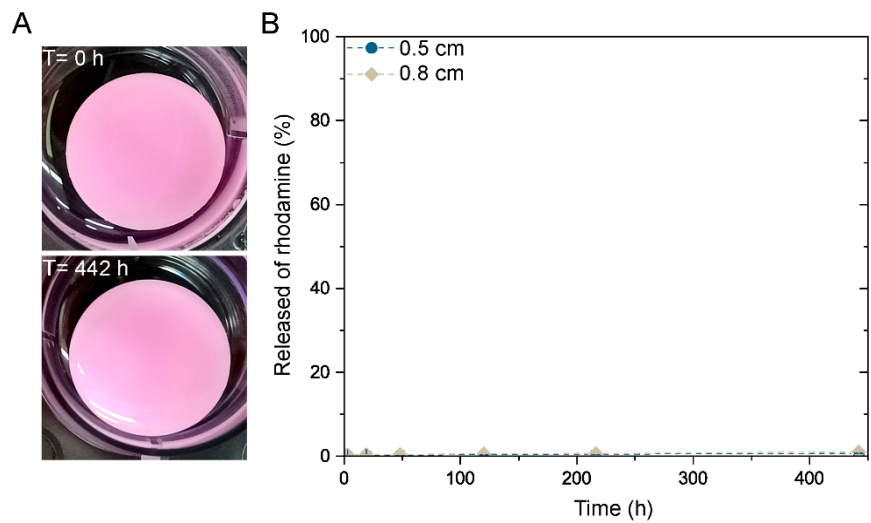

Figure S3. *In-vivo* cumulative release of PMN-rhodamine conjugate from hydrogel. (**A**) Pictures of PMN-hydrogel at $T_1$=0 h (upper) and $T_2$ = 442 h (lower), (**B**) Kinetic profile of PMN-rhodamine-conjugate released from hydrogel (0.5 cm or 0.8 cm thickness, diameter: 1.6 cm) upon incubation at physiological conditions (PBS, 37 °C).

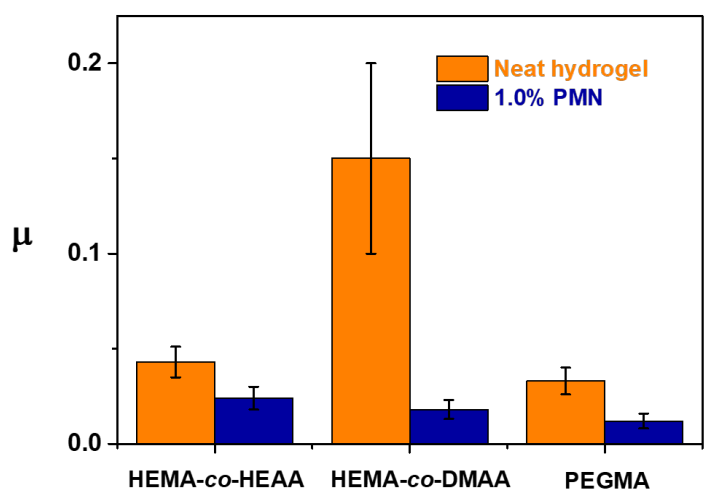

Figure S4. Sliding friction coefficients between a stainless-steel sphere and the different gels described above. Normal loads $F_n$ were (A) HEMA-co-HEAA: 10.0 N. (B) HEMA-co-DMAA: 10.0 N. (C) PEGMA: 2.0 N.

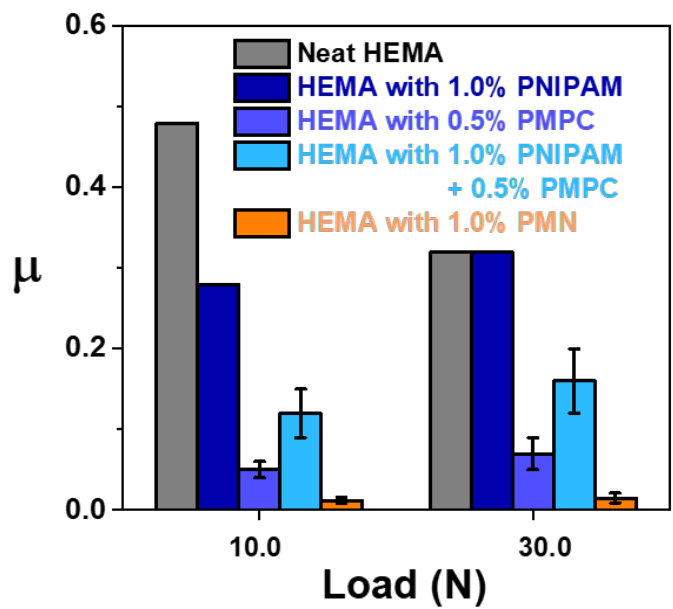

Figure S5. Variation of the coefficient of friction μ with different polymers in the hydrogel at different loads between a pHEMA hydrogel and a sliding steel sphere.

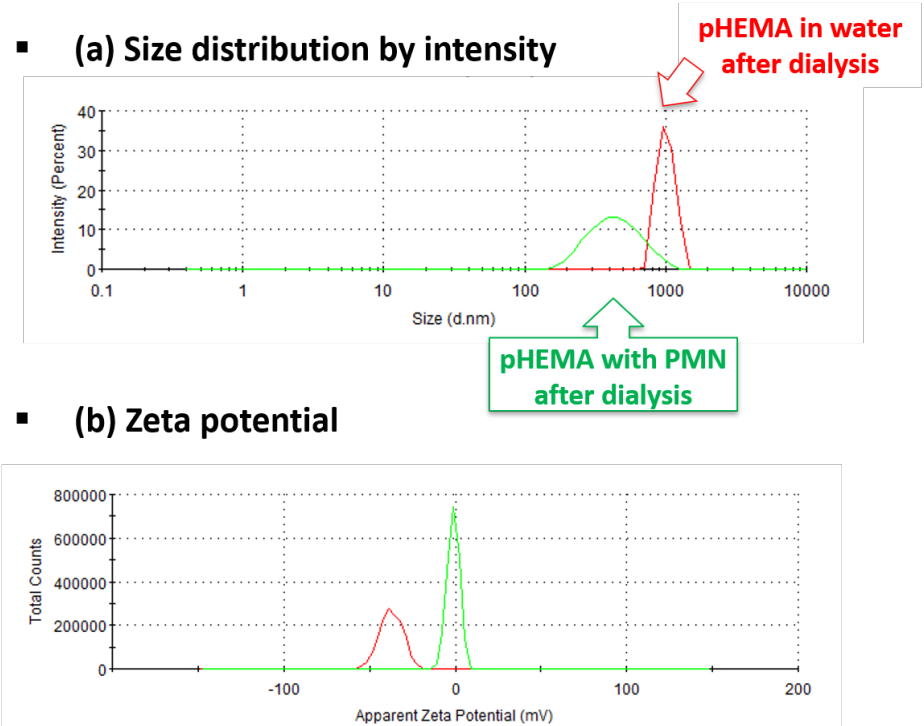

Figure S6. Characterization of pHEMA and pHEMA/PMN mixture in water by DLS (a) size distribution (b) zeta potential.

Table S1. Summary adsorption values measured from either SS and PE ball upon rubbing for 5 min with DMPC or PMN hydrogels. Values are presented as average and standard deviation from at least three separate samples

|  | SS ball | PE ball |
|---|---|---|
| **PMN** | Concentration: 2.20E-3 ± 1.888E-5 [mg/mL] <br> Area: 0.147 cm$^2$ | Concentration: 2.17E-3 ± 4.933E-5 [mg/mL] <br> Area: 0.1 cm$^2$ |
| **DMPC** | Concentration: 2.34E-6 ± 8.996E-7 [mg/mL] <br> Area: 0.166 cm$^2$ | Concentration: 9.60E-7 ± 3.44663E-7 [mg/mL] <br> Area: 0.18 cm$^2$ |

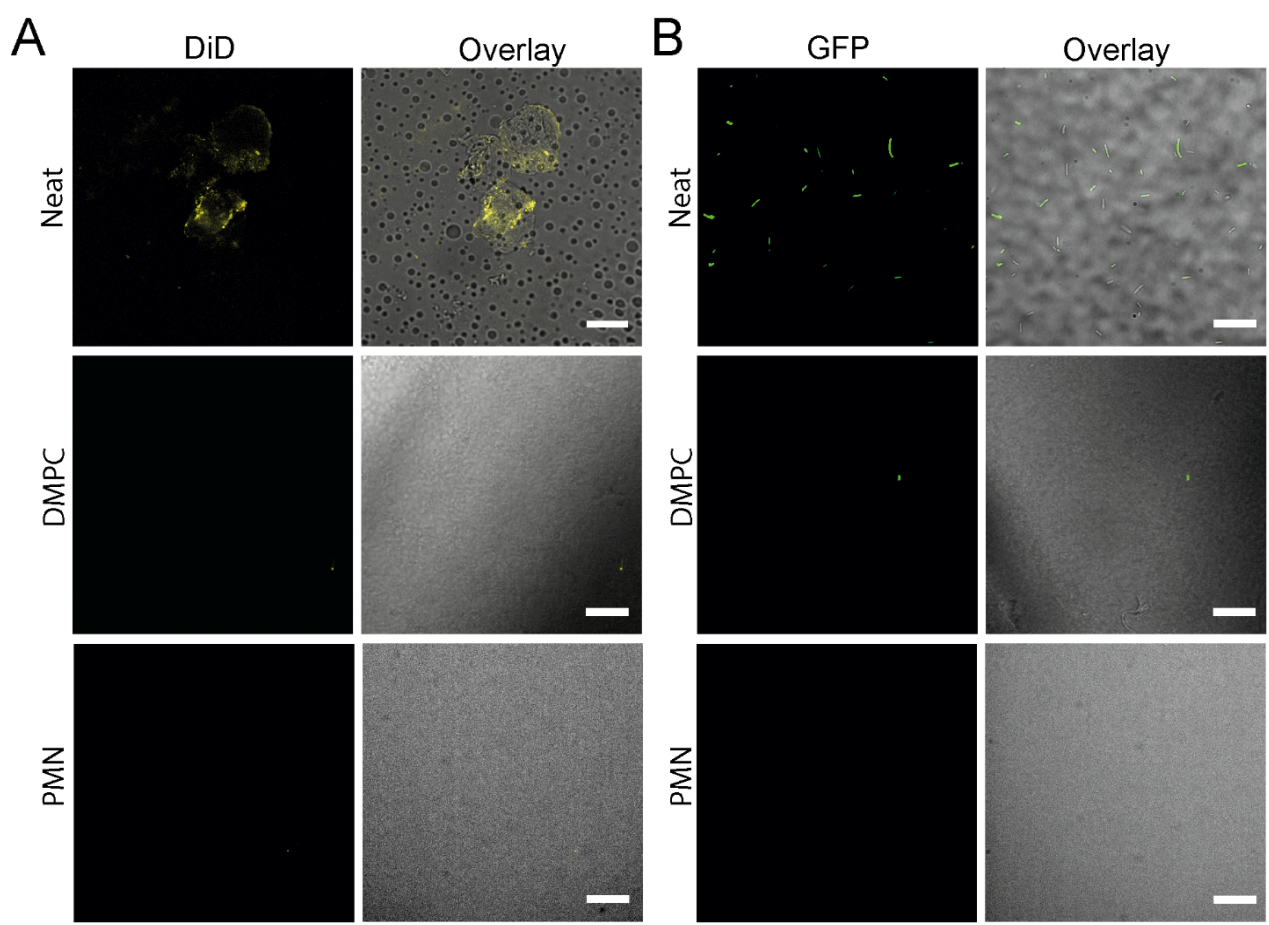

Figure S7. Confocal microscopy images of neat, DMPC and PMN hydrogels after 24 h incubation with (**A**) Vero E6 cells labeled with Vybrant DiD (yellow), and 12 h with (**B**) GFP-expressing E. Coli bacteria (green). Images are presented as respective fluorescence signals (yellow – DiD, and green-GFP) and overlays of brightfield and fluorescence staining. Scale bar, 20 μm.

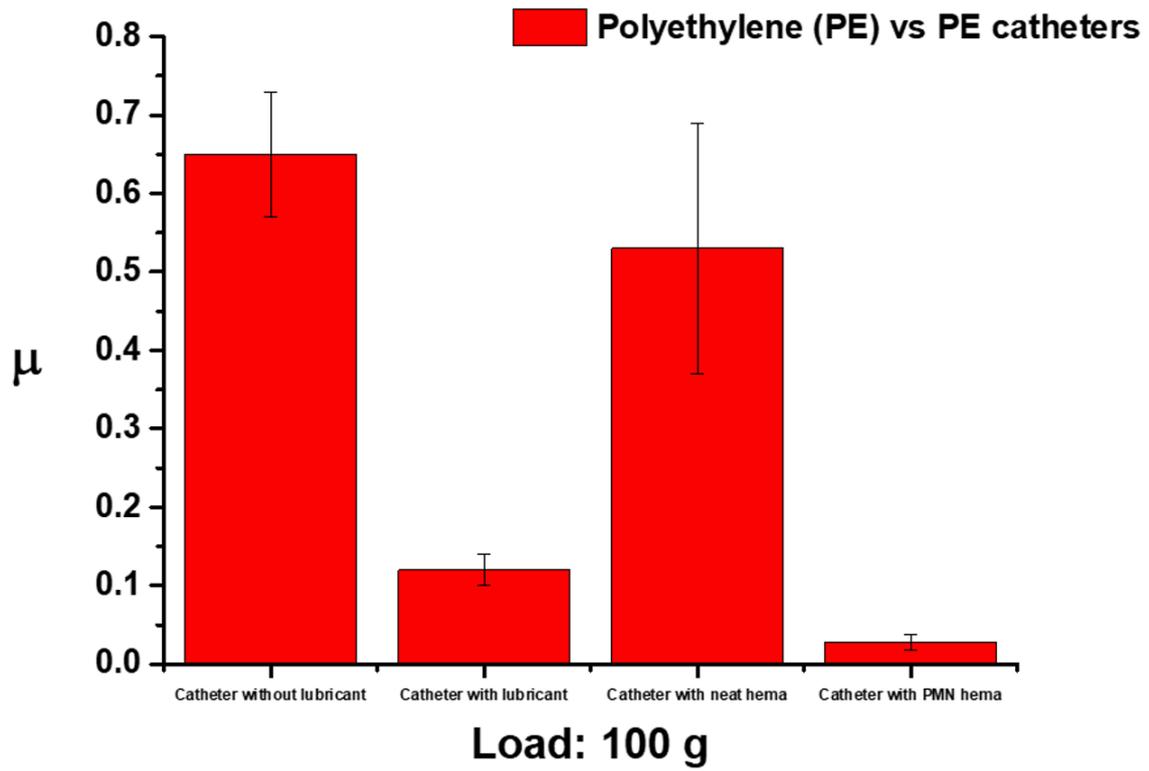

(a)

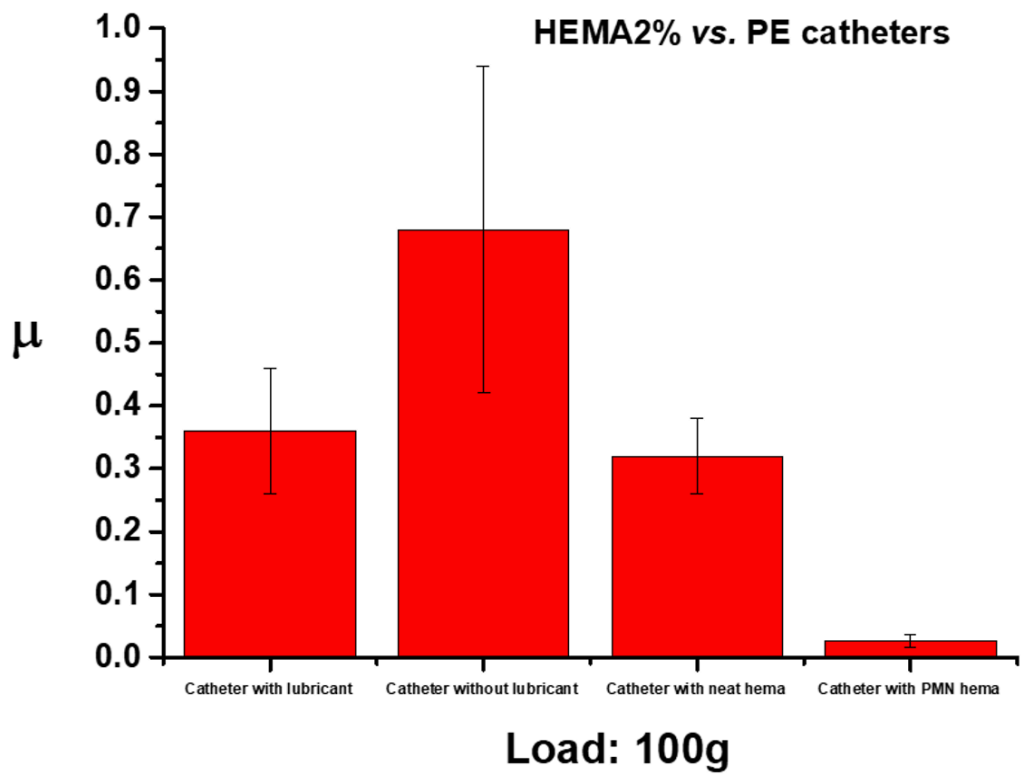

(b)

Fig. S8: Variation of the coefficient of friction μ at 100 g load between a commercial urethral catheter and a sliding polyethylene (hydrophobic) sphere (a) or a sliding pHEMA sphere (b). This shows that coating the catheter with PMN-containing pHEMA reduces the friction by up to 15-fold compared with the best available commercially-lubricated urethral catheter. PMN concentration in the gel was 2%, and the pHEMA was 2% x-linked.

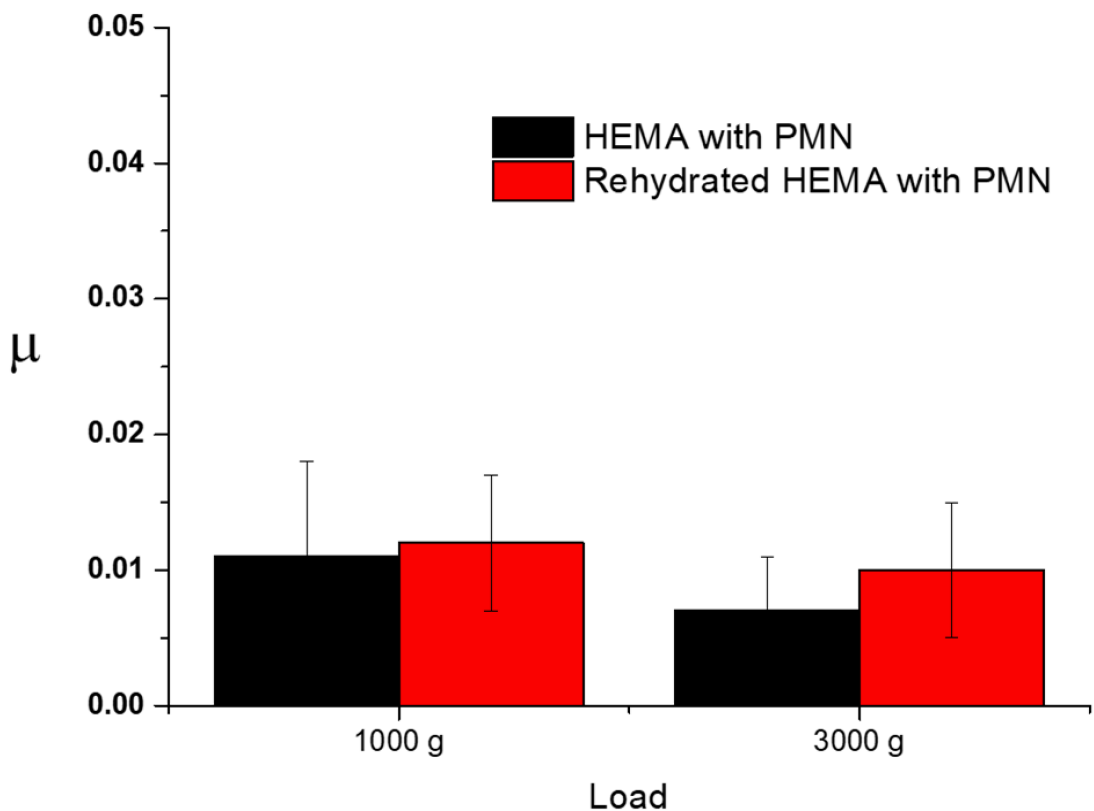

Fig. S9: Showing that following dehydration and rehydration (red columns) PMN-incorporating pHEMA gels fully recover their pre-dehydration lubricity (black columns). PMN concentration was 1%, and the pHEMA was 2% x-linked.